%% file: sample-sigconf-authordraft.tex
\documentclass[sigconf]{acmart}
\AtBeginDocument{%
  }

\setcopyright{acmlicensed}
\copyrightyear{2025}
\acmYear{2025}
\acmConference[Conference acronym 'XX]{Make sure to enter the correct
  conference title from your rights confirmation email}{June 03--05,
  2018}{Woodstock, NY}

\usepackage{amsmath,amssymb,amsfonts}
\usepackage{algorithmic}
\usepackage{graphicx}
\usepackage{textcomp}
\usepackage{xcolor}
\usepackage{hyperref}
\usepackage{listings}
\usepackage{subcaption}

\begin{document}

\include{macros}

\title{On the Challenges of Energy-Efficiency Analysis in HPC Systems: Evaluating Synthetic Benchmarks and Gromacs}

\author{Rafael Ravedutti Lucio Machado}
\affiliation{%
  \institution{Erlangen National High Performance Computing Center}
  \city{Erlangen}
  \state{Bavaria}
  \country{Germany}
}
\email{rafael.r.ravedutti@fau.de}

\author{Jan Eitzinger}
\affiliation{%
  \institution{Erlangen National High Performance Computing Center}
  \city{Erlangen}
  \state{Bavaria}
  \country{Germany}
}
\email{jan.eitzinger@fau.de}

\author{Georg Hager}
\affiliation{%
  \institution{Erlangen National High Performance Computing Center}
  \city{Erlangen}
  \state{Bavaria}
  \country{Germany}
}
\email{georg.hager@fau.de}

\author{Gerhard Wellein}
\affiliation{%
  \institution{Erlangen National High Performance Computing Center}
  \city{Erlangen}
  \state{Bavaria}
  \country{Germany}
}
\email{gerhard.wellein@fau.de}



\begin{abstract}
This paper discusses the challenges encountered when analyzing the energy efficiency of synthetic benchmarks and the Gromacs package on the Fritz and Alex HPC clusters.
Experiments were conducted using MPI parallelism on full sockets of Intel Ice Lake and Sapphire Rapids CPUs, as well as Nvidia A40 and A100 GPUs.
The metrics and measurements obtained with the Likwid and Nvidia profiling tools are presented, along with the results.
The challenges and pitfalls encountered during experimentation and analysis are revealed and discussed.
Best practices for future energy efficiency analysis studies are suggested.
\end{abstract}



\maketitle

\section{Introduction}

High-performance computing (HPC) systems are essential for advancing research and solving complex problems in many fields, such as molecular dynamics, biomedical simulations, and artificial intelligence.
However, the growing scale and complexity of modern HPC infrastructures has led to a significant increase in energy consumption, creating challenges in terms of cost, sustainability, and system reliability.
As energy becomes a limiting factor in system scalability and efficiency, it is increasingly important to conduct a comprehensive energy analysis.

This paper evaluates various aspects of conducting research experiments on HPC energy efficiency analysis.
It examines the power usage of synthetic benchmarks and real application codes to understand the practical differences between them and to demonstrate how simple benchmarks can be useful to understand the energy and performance characteristics of modern architectures.
Gromacs~\cite{lindahl2023gromacs}, a molecular dynamics package with a primary focus on life-science simulations, is used as a real-world application. This software is renowned for its high performance on HPC systems comprising both CPUs and GPUs.
The MD-Bench \cite{machado2023mdbench} prototyping harness for MD simulations is used as an additional benchmark.
It contains short-range force calculation kernel implementations that use the Cluster Pair data structure introduced by Gromacs to exploit optimal SIMD usage on modern CPUs and GPUs.

The experiments were conducted on two clusters: the \emph{Fritz} cluster, which consists of Intel Ice Lake (ICX) and Sapphire Rapids (SPR) nodes, and the \emph{Alex} cluster, which consists of nodes with Nvidia A40 and A100 GPUs.
The objectives are to (a) understand how different settings, such as power-capping limits and frequency, affect performance and energy efficiency in various simulation scenarios and hardware, and (b) share the insights gained from this systematic study and explain how performance metrics can be presented for a meaningful energy efficiency analysis.

Section \ref{sec:relatedwork} displays related work on the analysis of energy consumption and efficiency of HPC applications.
Section \ref{sec:background} provides a theoretical background on Gromacs and the performance monitoring tools utilized in this work.
Section \ref{sec:results} displays the challenges and results from our measurements, as well as a discussion on the pitfalls when doing research on performance and energy efficiency analysis on HPC systems.
Finally, Section \ref{sec:conclusion} provides the conclusions and outlines of this work.

\subsection{Related Work}
\label{sec:relatedwork}

As power constraints increasingly limit scalability in HPC systems, energy efficiency analysis at the application level has gained significant attention.
Although some research focuses on system-level optimizations and power management techniques, such as DVFS \cite{10.1145/3649329.3655956,11161883,10.1145/3710848.3710875} and power capping \cite{10.1002/cpe.4485,10.1007/978-3-030-43222-5_11,10820573}, with little focus on specific applications and hardware, many studies focus on understanding and improving the energy behavior of scientific applications themselves on specific machines.
Analysis at the application level for a specific hardware is crucial because energy consumption is determined by hardware characteristics, algorithmic choices, communication patterns, and computational intensity. 

Several studies have examined the energy-to-solution and performance trade-offs of representative scientific workloads, including molecular dynamics \cite{Savioli2025.10.08.681202}, computational fluid dynamics \cite{GODDEKE2013132}, and numerical linear algebra \cite{TAN2014559,doi:10.1177/1094342018792079}.
These studies often use measurement frameworks such as Likwid, Intel RAPL, or NVIDIA's NVML to collect power and performance metrics with high granularity.
By correlating these measurements with computational phases (i.e. force calculations, solver steps, or communication regions), researchers can identify the code sections that are responsible for the majority of energy consumption and then evaluate whether these can be improved.

In addition to the correlation and characterization of the energy and power consumption of one or more applications, there are studies that focus on providing analytic models to predict the performance and energy consumption of modern multi-core processors as a function of relevant parameters, such as frequencies and the number of active cores~\cite{Hager:2016,Hofmann:2018}.
These studies must be validated through synthetic benchmarks and real-world applications. Therefore, using a consistent methodology can help identify and overcome obstacles.

\section{Background and Theory}
\label{sec:background}

\subsection{Gromacs}

Gromacs is an open-source molecular dynamics simulation package designed to simulate systems at the atomic level, particularly biomolecular systems.
It implements advanced integrators and constraint solvers, as well as algorithms for treating long-range interactions, such as the Particle-Mesh Ewald method. It also provides data structures for efficiently computing short-range atomic forces.
Its popularity comes from its combination of numerical accuracy, high efficiency, and portability across a range of hardware platforms.

Performance optimization has been a central objective throughout the development of Gromacs.
The software exploits multiple levels of parallelism, including SIMD vectorization, OpenMP threading, MPI distribution, and GPU acceleration.
Its hybrid parallelization strategy enables efficient use of heterogeneous architectures ranging from desktop workstations to top supercomputers.
A distinctive feature of Gromacs parallelization is the separation of tasks between particle–particle (PP) computations, which dominate short-range force calculations, and PME computations for long-range electrostatics.
Dedicated MPI ranks can be assigned exclusively to PME, while others focus on PP, reducing the cost and overhead from communication (MPI reductions from the PME algorithm only need to be done for dedicated PME ranks) and improving scalability on large systems.
Combined with domain decomposition and dynamic load balancing, this strategy allows Gromacs to efficiently simulate systems with millions of atoms and to maintain high parallel efficiency on thousands of cores.

A key optimization in Gromacs for short-range non-bonded interactions is the Cluster Pair (CP) algorithm.
Instead of evaluating forces between individual particle pairs, particles are grouped into small spatial clusters (typically containing 4–8 atoms), and all interactions between two such clusters are computed in a single operation.
This design improves data locality, reduces branching, and enables efficient use of SIMD vectorization on modern CPUs as well as massive parallelism on GPUs.
In contrast, other MD packages rely primarily on Verlet neighbor-lists, where each atom maintains a list of neighbors within a cutoff distance (plus a buffer).
While Verlet lists are conceptually simple and robust, they lead to irregular memory access patterns and often result in less efficient use of SIMD units, since they require the usage of gather instructions.
The CP algorithm can be seen as an extension of the Verlet approach, since it also relies on neighbor searching, but it organizes the neighbor list at the cluster level rather than per atom.
Besides, it also stores atom properties (such as positions, velocities and forces) in an array-of-struct-of-arrays (AoSoA) layout, where data from atoms in the same cluster is contiguous in memory and can be loaded into the vector registers without gather instructions.
This reduces memory bandwidth requirements and also increases the arithmetic intensity by reusing loaded data for multiple pairwise computations.
However, there is a significant increase in the number of redundant computations in contrast to the Verlet lists because interactions are evaluated for all pair of atoms in the two clusters during the force calculation kernel. 
When combined with domain decomposition and dynamic load balancing, the CP algorithm contributes significantly to the strong performance and scalability of Gromacs compared to other MD packages.

As high-performance computing (HPC) systems scale toward exascale, energy efficiency has become as important as raw throughput.
MD simulations, particularly with codes like Gromacs, are computationally intensive and often run for extended time scales, leading to significant energy consumption.
Studying the energy efficiency of Gromacs provides valuable insights into how different computational phases (e.g., PP vs. PME kernels) and hardware resources (CPU vs. GPU, network communication) contribute to the overall energy footprint.
Such analyses help identify bottlenecks where energy use is disproportionate to performance gains, guiding both software optimization and hardware utilization strategies.
From a practical perspective, energy-to-solution is increasingly relevant for users and HPC centers alike, as it directly impacts operational costs and the sustainability of large-scale scientific computing.
Understanding and optimizing energy efficiency in Gromacs therefore supports not only scientific productivity but also the broader goals of green HPC and responsible resource usage.

\subsection{Synthetic Benchmarks}

\subsubsection{MD-Bench}
MD-Bench is a performance-oriented prototyping harness implemented in C99 that comprises the most essential MD steps to calculate trajectories in an atomic-scale system.
It contributes clean reference implementations of state-of-the-art MD optimization schemes from multiple community codes.
Since MD-Bench contains kernels that compute short-range atomic forces using the same Cluster Pair algorithm strategy from Gromacs, it can be utilized to approach performance and energy efficiency studies with Gromacs.
Kernel variants have been implemented for \texttt{full} (FN) and \texttt{half} (HN) neighbor lists.
The former stores each pair of interacting atoms twice (in both the $i$-$j$ and $j$-$i$ directions) and reduces forces for only one atom.
The latter case, which is also used by Gromacs, stores atom pair interactions only once (in the $i$-$j$ direction) when the forces are symmetric.
Using half-neighbor lists reduces computation and data loads by roughly a factor of two but introduces gather and scatter operations to reduce forces in the $j$-atom; it also requires atomic operations when using shared-memory parallelism.


\subsubsection{Stream}

\lstset{
  basicstyle=\ttfamily\footnotesize,
  keywordstyle=\color{blue}\bfseries,
  commentstyle=\color{gray}\itshape,
  stringstyle=\color{teal},
  numbers=left,
  numberstyle=\tiny\color{gray},
  stepnumber=1,
  numbersep=8pt,
  tabsize=4,
  showstringspaces=false,
  breaklines=true,
  frame=single,
  captionpos=b
}

The \textbf{STREAM} benchmark family measures sustainable memory bandwidth and corresponding computation rates for simple vector kernels that are representative of many memory-bound scientific applications.
In this work, we use the triad kernel, which computes $A[i] = B[i] + \alpha \cdot C[i]$ for single-precision floating-point arrays $A$, $B$, $C$ of size $N$.

\subsubsection{GEMM}

The \textbf{GEMM} (General Matrix Multiply) benchmark computes dense matrix multiplications of the form \(C = \alpha A B + \beta C\).
There are optimized implementations of the benchmark in libraries such as \texttt{cuBLAS} and \texttt{Intel MKL}, and these are well-known representative scenarios for assessing instruction throughput bound applications on both CPUs and GPUs.
The benchmarks are usually referred to as \textbf{SGEMM} and \textbf{DGEMM}, where the former works on single 32-bit floating-point precision and the latter on double 64-bit floating-point precision.
Although modern GPU implementations can use Tensor Cores to achieve higher performance throughput, it was decided not to use them, as they are tailored for matrix multiplication workloads, unlike Gromacs.
Instead, we stressed the CUDA cores.







\section{Materials and Methods}
\label{sec:materials}


\begin{table*}[htb]
  \centering
  \begin{tabular}{c|c|c|c|c}
    Model & Xeon Platinum 8470 & Xeon Platinum 8360Y & A40 & A100 \\
    Micro-architecture & Sapphire Rapids (SPR) & Ice Lake server (ICX) & Ampere & Ampere \\
    \hline
    Base frequency & 2.0\,\GHZ & 2.4\,\GHZ & 1.065\,\GHZ & 1.305\,\GHZ  \\
    Boost clock & 3.8\,\GHZ & 3.4\,\GHZ & 1.410\,\GHZ & 1.740\,\GHZ  \\
    Memory clock & --- & --- & 1.512\,\GHZ & 1.812\,\GHZ  \\
    Cores/SMs per chip & 52 & 36 & 84 & 108 \\
    Latest SIMD extension& AVX512 & AVX512 & --- & ---  \\
    TDP & 350W & 250W & 300W & 400W  \\
    L1D capacity & $52 \times 48$\,\KiB & $36 \times 48$\,\KiB & $84 \times 128$\,\KiB & $108 \times 192$\,\KiB  \\
    L2 capacity & $52 \times 2$\,\MiB & $36 \times 1.25$\,\MiB & $6$\,\MiB & $80$\,\MiB  \\
    L3 capacity & 105\,\MiB & 54\,\MiB & --- & ---  \\
    Memory configuration & 6 ch. DDR4-2933 & 8 ch. DDR4-3200 & --- & --- \\
    Memory size  & --- & --- & 48\,\GiB & 80\,\GiB  \\
    Peak FP32 Perf. (TFLOP/s) & 6.656 & 5.529 & 37.42 & 19.49 \\
    Peak Bandwidth  & --- & --- & 695.8 GB/s & 1.94 TB/s \\
  \end{tabular}
  \caption{Specifications for testbed processors used.}
  \label{tab:testbed}
\end{table*}


\autoref{tab:testbed} lists the CPU and GPU processors used in this work and displays their relevant settings.
Experiments were conducted on two Intel CPUs available in the Fritz cluster: an Intel Sapphire Rapids system and an Intel Ice Lake system.
For GPU experiments, two different GPU models available in our Alex cluster were used: an A40 and an A100, both featuring the Ampere architecture.
For each node in the Alex cluster, 8 units of the respective GPUs are present together with 2 AMD EPYC 7713 ("Milan", "Zen3") CPUs, each with 2 x 64 cores at 2.0 GHz base clock frequency.
Each node in the Alex cluster has 8 of the respective GPUs together with 2 AMD EPYC 7713 ("Milan", "Zen3") CPUs, each with 2 x 64 cores at a base clock frequency of 2.0 GHz.
While some of the Gromacs phases run on the CPU in a hybrid setting, the heavy computational workloads, such as the PP and PME kernels, are offloaded to the GPU. 
Therefore, our analysis is  performed exclusively on the GPUs in this work.

The LIKWID software \cite{LIKWID:2012} was utilized to pin tasks for specific cores and measure HPM counter metrics.
The configuration of the CPU frequency settings is facilitated by the SLURM option \emph{--cpu-freq} in conjunction with the \emph{performance} governor, a configuration that ensures the continuous operation of the CPU at its maximum frequency capacity.
In the context of GPUs, the \emph{nvidia-smi} tool is employed to modify the frequency and power capping settings. This is achieved by utilizing the \emph{set-applications-clock} and \emph{power-limit} options, respectively.



The Gromacs 2023.4 version was used and built from the repository to include the proper Likwid instrumentation markers in the code.
For this, the ICX 2023.2.0.20230721 compiler was used with the Intel MPI 2021.7.0 and Intel MKL 2023.2.0 modules.
In the GPU systems, the module \emph{gromacs/2023.3-gcc11.2.0-openmpi-mkl-cuda}, which indicates that Gromacs 2023.3 compiled with the GCC 11.2.0 compiler was used.
Although many other Gromacs test cases were evaluated, this study covers the ones that showed the most interesting results and did enough work to avoid major impacts caused by the overhead.
The benchmarks were selected from a list of representative cases\footnote{\url{https://hpc.fau.de/2022/02/10/gromacs-performance-on-different-gpu-types/}}:
\begin{itemize}
    \item \textbf{B4:} a protein in explicit water (170,320 atoms)
    \item \textbf{B5:} a protein membrane channel with explicit water (615,924 atoms)
\end{itemize}
Both benchmarks were executed for a duration of 200,000 time steps, with the majority of cases continuing for more than 10 minutes to ensure the accurate rendering of energy and power consumption behaviors.

In the context of MD-Bench, the standard Copper Lattice simulation is employed, encompassing  $128^3$ unit cells (8.388.608 atoms), a cutoff radius of $4.0$ and a "skin" of $0.5$ over a duration of 10,000 time steps.
Since the MD-Bench GPU implementation still has nonoptimal performance for building the neighbor-lists, MD-Bench is only used on CPU systems.



\section{Challenges and best practices}
\label{sec:results}

\subsection{Meaningful energy and performance profiling}

\begin{table*}[htb]
\centering
\begin{tabular}{l|c|c|c|c}
Kernel/Architecture & PP/SPR & PP/ICX36 & PME/SPR & PME/ICX \\
\hline
Runtime@base\_freq (s) & 3470.811 & 6439.544 & 1481.100 & 3837.322 \\
Number of ranks & 40 & 30 & 12 & 6 \\
Wait time (s) & --- & --- & 2332.901 & 3415.535 \\
Giga cycles (\%) & 70.0 & 9.0 + 14.1 (wait) & 74.0 & 8.8 + 7.8 (wait) \\
All Instr & $6.732 \times 10^{14}$ & $6.717 \times 10^{14}$ & $1.208 \times 10^{14}$ & $1.005 \times 10^{14}$ \\
FP32 Instr. AVX512 & $4.468 \times 10^{14}$ & $4.440 \times 10^{14}$ & $9.328 \times 10^{12}$ & $9.328 \times 10^{12}$ \\
FP32 Instr. AVX2 & $1.676 \times 10^{13}$ & $1.664 \times 10^{13}$ & $1.200 \times 10^{5}$ & $6.000 \times 10^{4}$ \\
FP32 Instr. SSE & $1.782 \times 10^{12}$ & $1.785 \times 10^{12}$ & $8.618 \times 10^{12}$ & $8.618 \times 10^{12}$ \\
FP32 Instr. Scalar & $2.130 \times 10^{12}$ & $2.121 \times 10^{12}$ & $1.051 \times 10^{13}$ & $9.903 \times 10^{12}$ \\
Arith. Instr. Ratio (\%) & 69.43 & 69.15 & 23.55 & 27.72 \\
Estimated TFLOPS/s & 2.1 & 1.125 & 0.131 & 0.05 \\
Theoretical peak@nranks & 5.12 & 4.6 & 1.53 & 0.92\\
\end{tabular}
\caption{Performance and instruction metrics for PP and PME configurations on SPR and ICX systems.}
\label{tab:pp_pme_metrics}
\end{table*}

Without the ability to accurately measure hardware metrics, energy studies are impossible.
Key issues include which metrics are available, the scope of the measurements, the overhead introduced, and the minimum sampling rate at which metric values are updated at hardware level.

With Gromacs the process of profiling can result in the introduction of a substantial overhead, primarily due to the fine-grained nature of its computational phases and the significantly short runtime of each simulation timestep.
For instance, when utilizing the two SPR sockets (104 threads) at a fixed frequency of 2.0\,GHz, the runtime of the non-bonded force computation kernel per timestep is approximately $10^{-3}$s.
Due to the short execution time of each step, even minor overhead can significantly impact the accuracy of measurements performed at the kernel level.


For this reason, the use of Likwid instrumentation markers is limited in this situation.
These markers trigger system calls, which significantly increase runtime and alter the overall characteristics of the code, since more time is spent on instrumentation than on the actual measured code.
In addition, the RAPL counters, which are used most often, usually have a low update frequency, which further influences the results for very small regions.

It is still possible to use instrumentation for metrics that are not sensitive to time or other types of profiling noise.
Examples include total and arithmetic instructions.
\autoref{tab:pp_pme_metrics} displays data on single-precision floating-point arithmetic instructions for the Gromacs test-case B5, which was collected using Likwid instrumentation on the PP and PME kernels.
It is even possible to estimate the performance in {Tflop/s} for each region by using the runtime from a plain, non-instrumented run.
It is evident that PP kernels utilize significantly more compute units than PME phases. There are a substantial number of instructions for all SIMD sets, indicating the reduction of values within SIMD registers, such as the computed forces. Approximately 70\% of the instructions are for arithmetic computation.
The PME ranks spent a significant amount of time and cycles waiting for other ranks to complete, and only about 20-30\% of the instructions are arithmetic.
This is already valuable information which can be obtained via instrumentation despite the limitations described.

An alternative to instrumentation is to use the Linux \emph{perf\_events} tool and perform sampling.
This should provide less overhead since \emph{perf\_events} operates directly in the kernel and sampling can be adjusted to a frequency that leads to less overhead.
However, issues were encountered when using this approach with Gromacs:

\begin{itemize}
    \item \textbf{Overhead still significant:} Despite utilizing a sampling frequency as low as 99 Hz, a 20-30\% overhead is observed in the context of floating-point arithmetic and clock-related events.
    \item \textbf{Large volume of data produced:} For a single execution of a modest Gromacs benchmark, an aggregate of 28–40 GB of data is obtained from the \emph{perf\_events} trace. This issue can be resolved by adjusting specific options and omitting sampling of the entire simulation. However, this remains a concern when executing multiple benchmarks and diverse scenarios.
    \item \textbf{Sampling of energy events not possible:} Energy events are counted per socket, hence when performing function-level profiling at the hardware threads level, the kernel cannot refer back from the socket energy to the contribution of each function call by the threads; This means that it is not possible to sample functions when events are counted per socket when using \emph{perf\_events}.
    \item \textbf{Inaccuracies with coarser, low-frequency sampling:} Since Gromacs computational phases run for a very short amount of time, sampling at larger frequencies can provide misleading results as the number of samples per function execution is not significant enough; Even sampling at a $99Hz$ frequency, we already face significant overhead, so determining a frequency that can at the same time provide negligible overhead and provide statistically sound results has shown to be an obstacle for our case.
\end{itemize}

Due to the challenges and pitfalls of finer-granularity profiling, end-to-end measurements were chosen as an alternative for power, energy, and clock frequency metrics.
While certain assumptions can be made, it is evident that it is not possible to determine the performance and energy characteristics of each individual kernel based on the counted events.
Although this paper does not focus on HPM overhead evaluation or the capabilities and user-friendliness of HPM facilities, it is important to acknowledge that the situation has deteriorated considerably in recent years.
Fine-grained, end-to-end measurements are impossible due to the overhead introduced by recent hardware security mitigations in operating system kernels and processor firmware.
HPM implementations on the CPU side are becoming more complicated, requiring more accesses and therefore system calls, which further contributes to large overheads.
Finally, some chip vendors still do not regard HPM metrics as important core product features.
For example, counters such as the Uncore clock ticks on Sapphire Rapids disappear between CPU generations, and no responsibility is taken for counter accuracy in general.
As a workaround clock tick counters from the interfaces between the mesh and the Intel UPI link layer (M3UPI) and between the mesh and each IIO stack (M2PCIE) can be used on Sapphire Rapids to estimate the Uncore frequency.



\begin{figure*}[htb]
  \centering
  \begin{subfigure}{0.33\textwidth}
    \includegraphics[width=\linewidth]{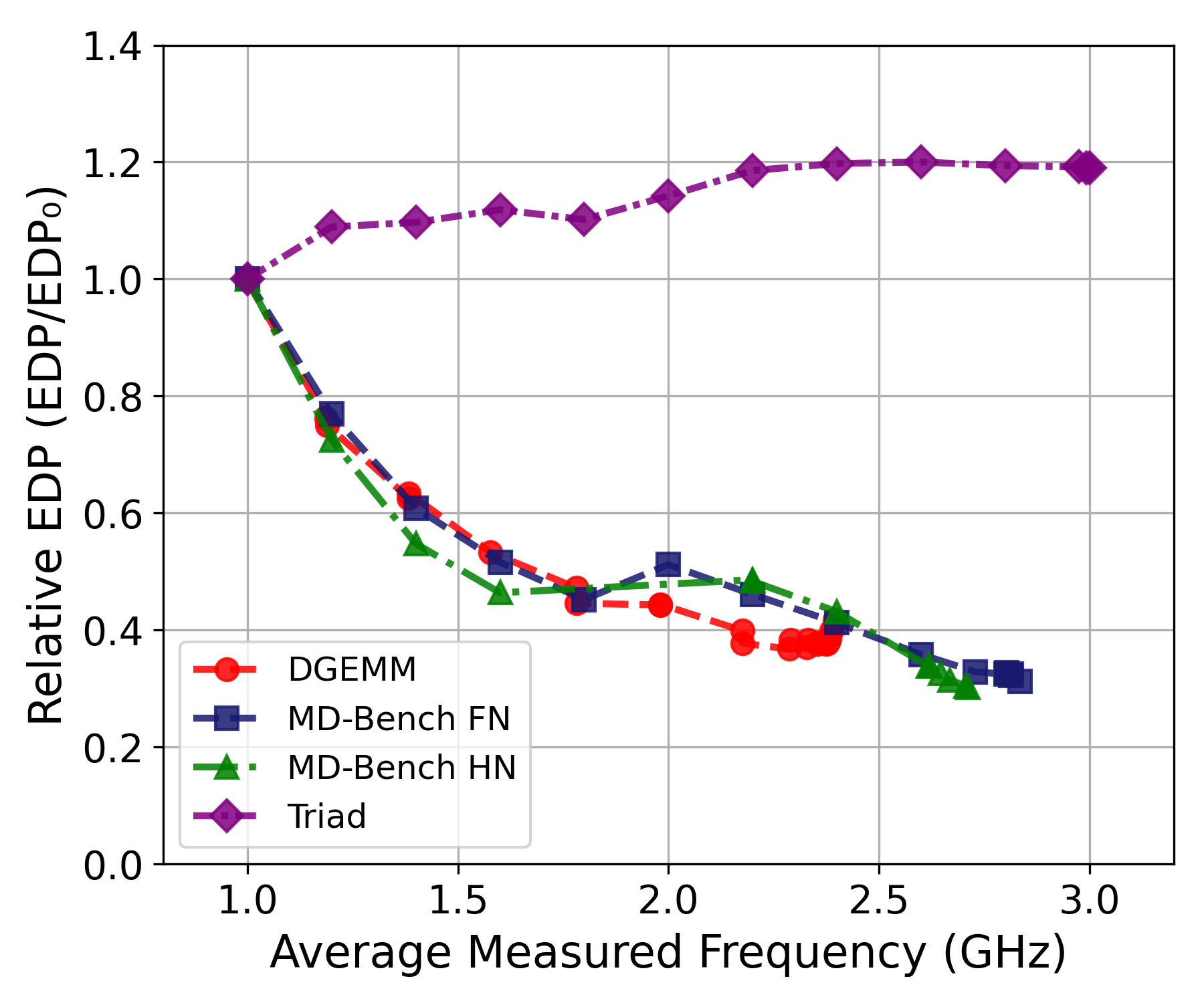}
    \caption{Synthetic benchmarks}
    \label{fig:synth_cpu1}
  \end{subfigure}
  \begin{subfigure}{0.33\textwidth}
    \includegraphics[width=\linewidth]{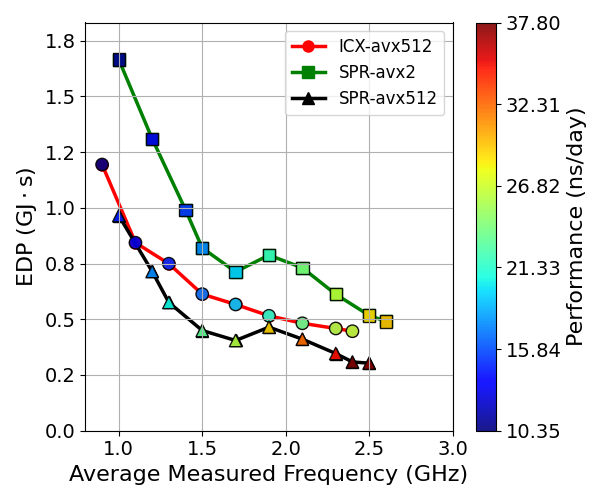}
    \caption{Gromacs B4}
    \label{fig:b4_edpxfreq}
  \end{subfigure}
  \begin{subfigure}{0.33\textwidth}
    \includegraphics[width=\linewidth]{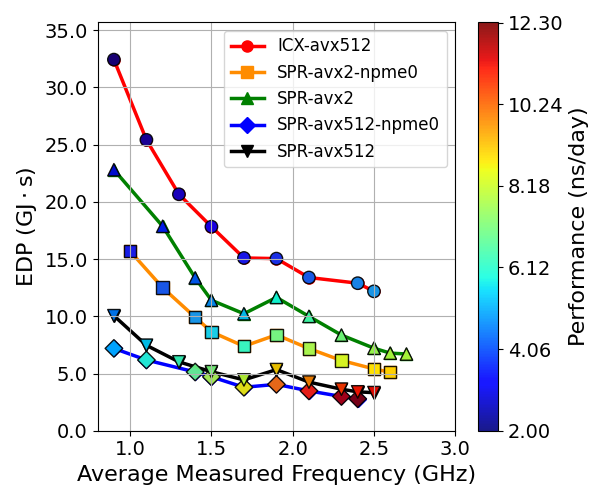}
    \caption{Gromacs B5}
    \label{fig:b5_edpxfreq}
  \end{subfigure}
  \caption{EDP versus measured frequencies for synthetic benchmarks on Sapphire Rapids and Gromacs (both Sapphire Rapids and Ice Lake). For Figure (a) with the synthetic benchmarks, the EDP is normalized. The marker colors represent the performance obtained on the Gromacs benchmarks in $ns/day$.}
  \label{fig:edp_comparison}
  \Description{EDP versus measured frequencies for synthetic benchmarks on Sapphire Rapids and Gromacs (both Sapphire Rapids and Ice Lake). For Figure (a) with the synthetic benchmarks, the EDP is normalized. The marker colors represent the performance obtained on the Gromacs benchmarks in $ns/day$.}
\end{figure*}

Another interesting approach to handle these limitations is the use of specific benchmark codes that resemble the original application.
One example for MD workloads is MD-Bench, which contains several implementations of such kernels using different strategies.
Synthetic benchmarks like DGEMM and Triad can also be used to understand how the runtime and energy behavior of the target machine behaves in situations where specific resources such as the compute units or the memory bandwidth are fully utilized.
\autoref{fig:edp_comparison} displays the Energy-Delay-Product (EDP) versus the measured frequency for synthetic benchmarks and for different Gromacs test-cases.
The best EDP case for Triad (memory-bound) occurs at the minimum frequency. For DGEMM, MD-Bench, and Gromacs, the EDP improves as the core frequency increases.
The only exceptions occur when the base frequency is reached, at which point a leap in EDP can be seen.

As expected, the EDP behavior is similar for both MD-Bench and Gromacs.
This reinforces that MD-Bench can be used for energy efficiency studies to have a better understanding of how MD                                                                                       codes behave with respect to performance and energy consumption; at the same time, it presents options that allow one to control the runtime execution for each phase and other parameters that affect its behavior (e.g. FN vs HN).
In such situations, it is important to investigate how transferable the knowledge obtained with the proxy application is.
This can be done by evaluating it against different scenarios with the real production code.

\subsection{Application specific influences on the example of Gromacs}

\begin{figure*}[tb]
  \centering
  \begin{subfigure}{0.33\textwidth}
    \includegraphics[width=\linewidth]{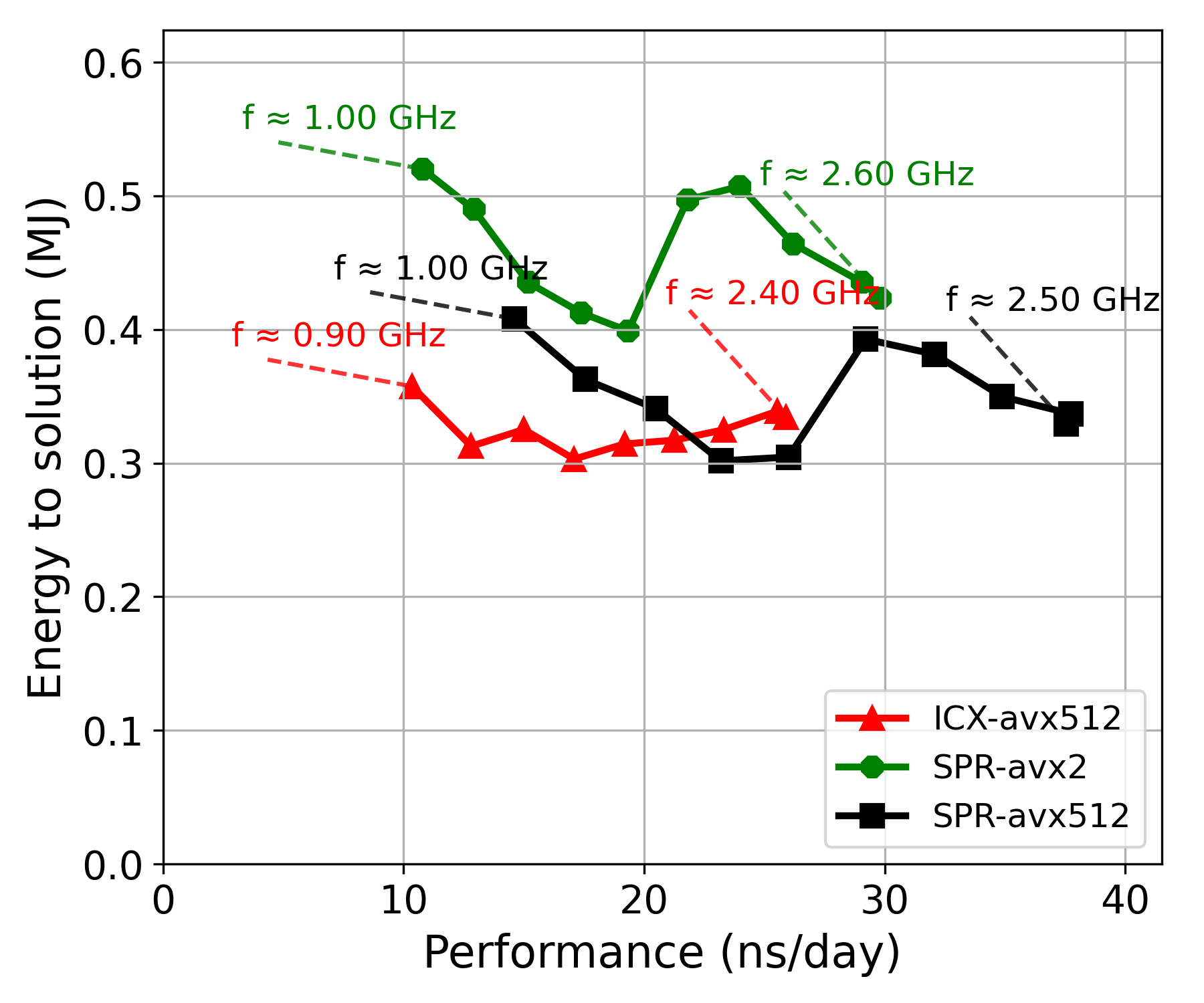}
    \caption{Gromacs B4 on CPU}
    \label{fig:b4_perfxenergy_cpu}
  \end{subfigure}
  \begin{subfigure}{0.33\textwidth}
    \includegraphics[width=\linewidth]{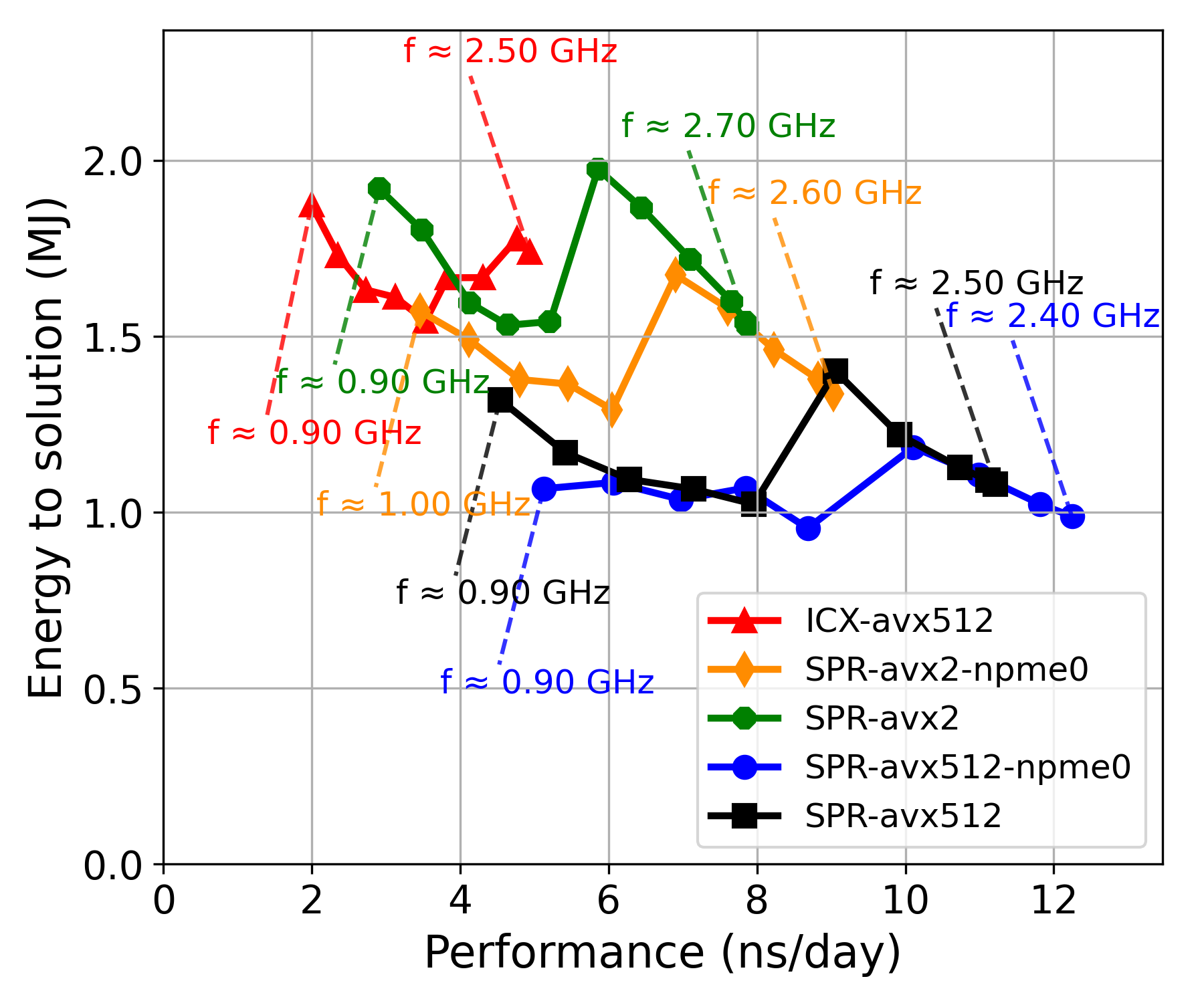}
    \caption{Gromacs B5 on CPU}
    \label{fig:b5_perfxenergy_cpu}
  \end{subfigure}
  \begin{subfigure}{0.33\textwidth}
    \includegraphics[width=\linewidth]{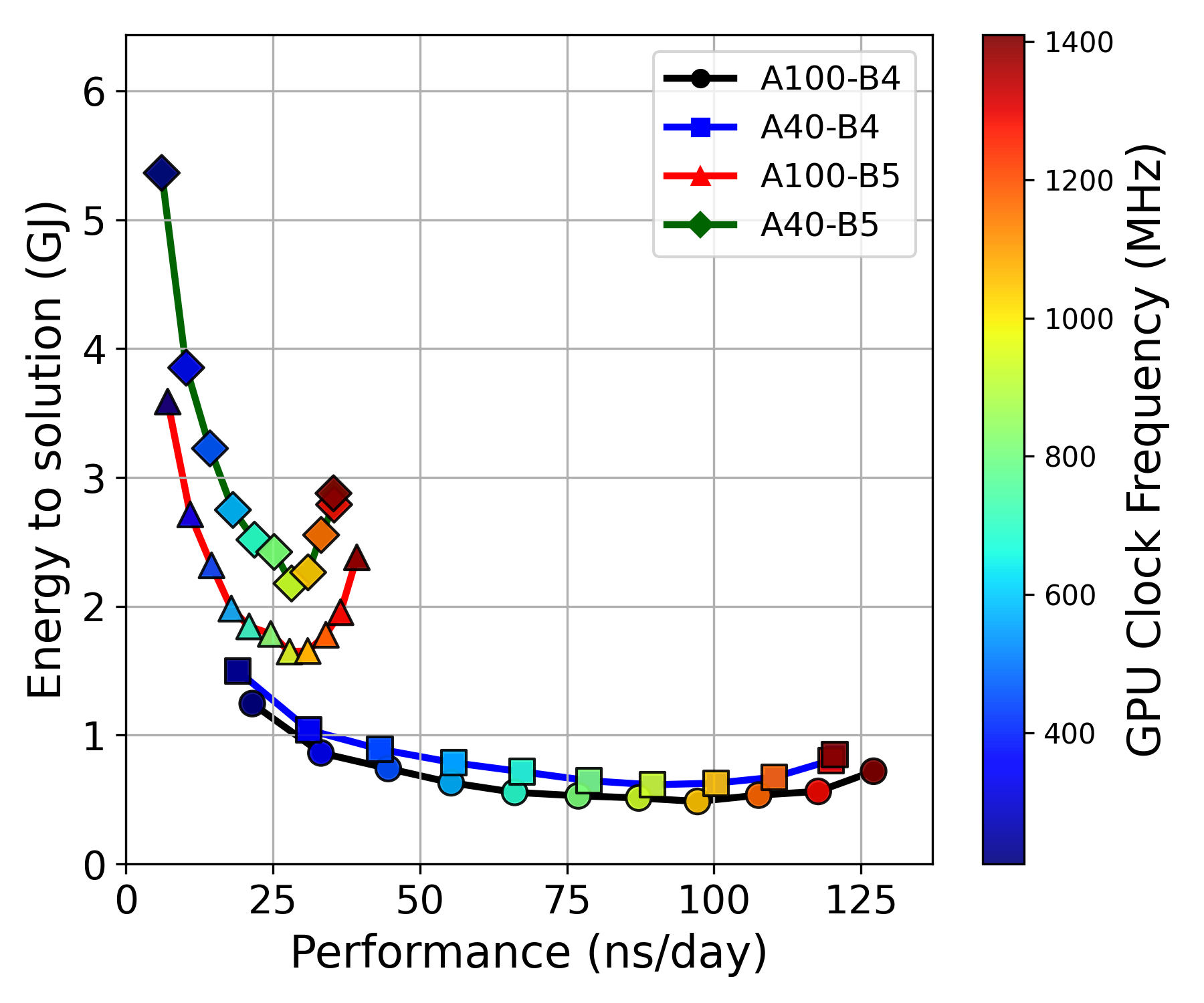}
    \caption{Gromacs B4 and B5 on GPU}
    \label{fig:perfxenergy_gpu}
  \end{subfigure}
  \caption{Z-plot showing the energy-to-solution versus performance when running different Gromacs benchmarks at different frequencies (measured) on both CPUs and GPUs. The marker colors on the Figure (c) represent the GPU graphics frequency setting.}
  \label{fig:perfxenergy_all}
   \Description{Z-plot showing the energy-to-solution versus performance when running different Gromacs benchmarks at different frequencies (measured) on both CPUs and GPUs.}
\end{figure*}

The performance and energy consumption of Gromacs are also affected by the choice of PP/PME rank decomposition.
In our case, a set of MPI ranks, which correspond to specific CPU cores, is selected to compute long-range interactions using the PME algorithm, while the rest are used to compute short-range or particle-particle interactions.
If not specified via parameters, Gromacs sets the number of ranks reserved for PME computations based on heuristics, hardware, and simulation settings.
Not all settings are valid for the number of PME ranks.
For example, on test-case B4 it is not possible to use all ranks for PP and PME (i.e. setting \texttt{npme} to zero) on the full SPR socket due to the lack of a valid domain decomposition with 52 ranks.

Allocating too few PME ranks can cause long-range computations to become a bottleneck, which limits parallel efficiency and increases time to solution.
Conversely, assigning too many PME ranks reduces the number of ranks available for short-range computations and increases communication overhead between the PP and PME domains.
Additionally, PME computation requires global synchronization between ranks, meaning that the communication cost is proportional to the number of ranks assigned to PME.

These decisions not only affect execution time but also alter the power distribution across compute nodes.
For instance, heavily loaded PP ranks often run at higher frequency and power levels, whereas PME ranks are more communication-bound and power-efficient.
Consequently, when conducting performance or energy studies, it is essential to understand the types of parallelism employed in the target applications and to analyze specific ranks or threads to see whether their behavior varies significantly.
In this work, this is only addressed for the CPU experiments since no multi-GPU experiments were performed, and both PP and PME computations were offloaded to the single GPU.

In \autoref{fig:b4_edpxfreq} and \autoref{fig:b5_edpxfreq} it is possible to observe that the different benchmarks exhibit different characteristics with respect to EDP on different architectures.
For test-case B4, the best EDP is achieved with AVX512 instructions on both SPR and ICX.
However, for B5, all SPR runs (even those using AVX2 instructions) have a lower EDP than ICX runs using AVX512 instructions.
Allocating all ranks to compute both PP and PME provides a lower EDP for B5, but as previously mentioned, this setting is not possible for B4 on SPR.
It is worth noting that the outcome may differ when running with more MPI ranks and resources.
Allocating more ranks for PME can lead to higher communication costs, which can worsen the EDP due to the impact of global reductions on performance.

For a more detailed overview of EDP behavior, \autoref{fig:b4_perfxenergy_cpu} and \autoref{fig:b5_perfxenergy_cpu} display the energy-to-solution versus performance for the same cases when running at different frequencies.
For test-case B4, the ICX version has worse performance at the highest frequencies than the AVX2 run on SPR.
However, this is compensated for by the lower energy consumption.
All AVX512 SPR cases perform better and consume less energy than their AVX2 counterparts. The effect is the same when all ranks are allocated to compute both the PP and PME phases.

In \autoref{fig:perfxenergy_gpu} the results for both benchmarks on the A40 and A100 GPUs.
The behavior matches the expectations \cite{Hofmann:2018} where a minimum energy-to-solution is achieved when running at about $1.0\,\GHZ$ before it starts to increase at higher frequencies.
Note that although the B4 curve looks more pronounced, this is actually caused by the energy-to-solution and performance scale.
The general behavior is similar in both cases.
Even though the A100 GPU has lower FP32 peak performance than the A40 GPU, Gromacs still runs on the A100 with slightly better performance.
This means that other factors, such as higher bandwidth and performance for other data types, compensate for the fewer FP32 units available.

\begin{figure*}[htb]
  \centering
  \begin{subfigure}{0.33\textwidth}
    \includegraphics[width=\linewidth]{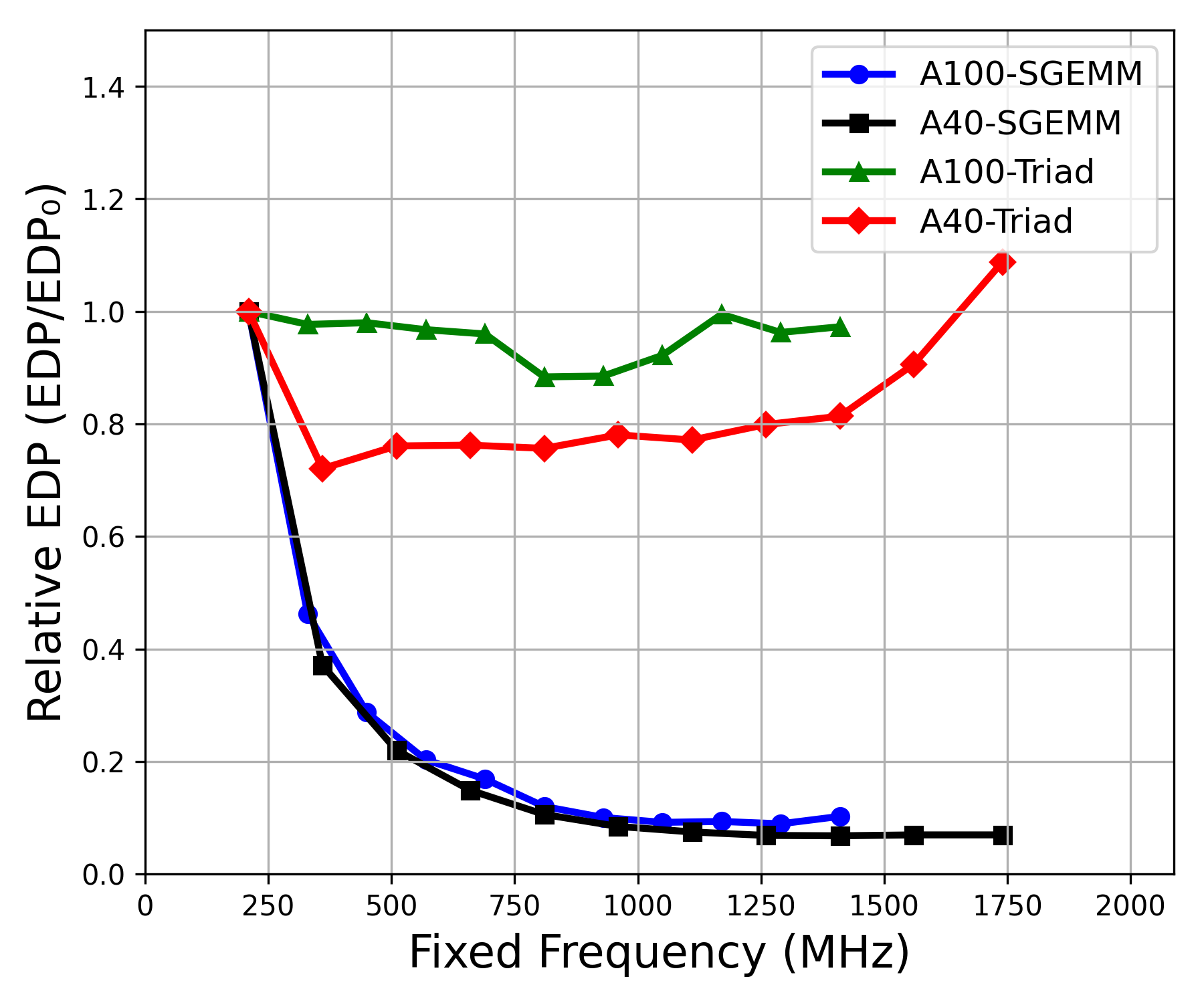}
    \caption{Synthetic benchmarks}
    \label{fig:edpxfreq_synth_gpu}
  \end{subfigure}
  \begin{subfigure}{0.33\textwidth}
    \includegraphics[width=\linewidth]{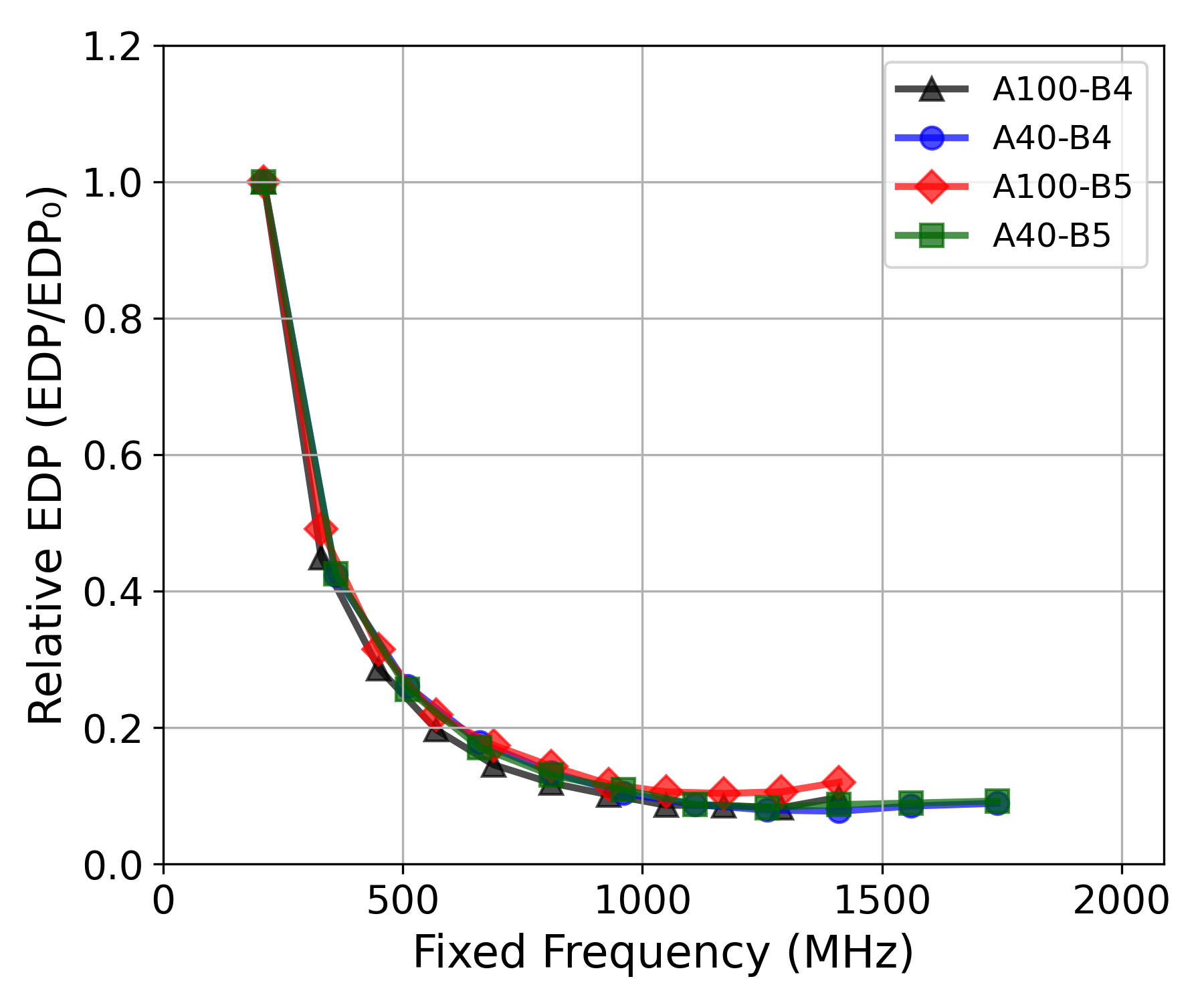}
    \caption{Gromacs EDP vs Frequency settings}
    \label{fig:edpxfreq_gpu}
  \end{subfigure}
  \begin{subfigure}{0.33\textwidth}
    \includegraphics[width=\linewidth]{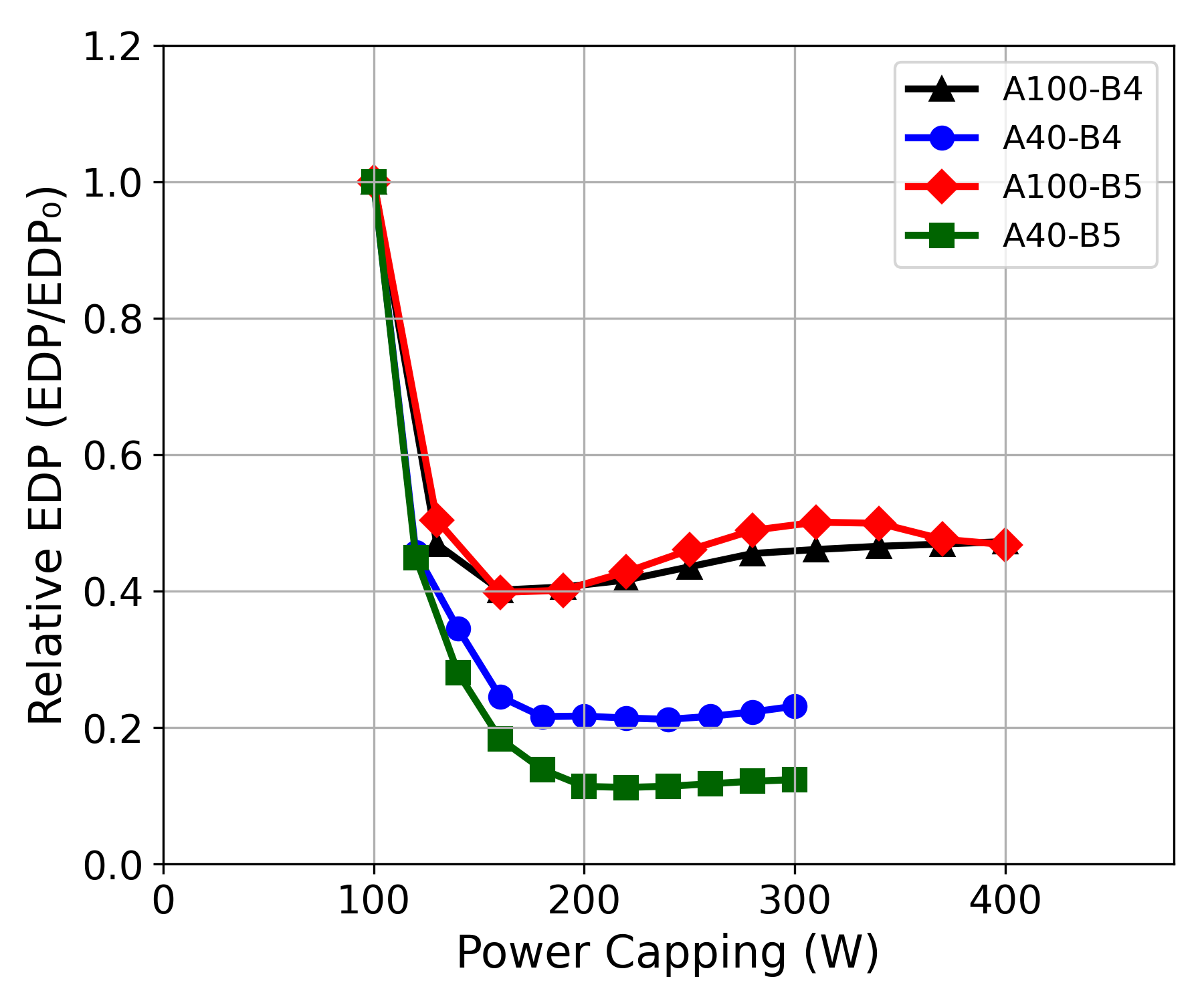}
    \caption{Gromacs EDP vs Powercap settings}
    \label{fig:edpxpowercap_gpu}
  \end{subfigure}
  \caption{EDP versus frequencies and powercap settings for synthetic benchmarks and Gromacs on A40 and A100 GPUs. For Figure (a) with the synthetic benchmarks, the EDP is normalized.}
  \label{fig:edpxfreqs_gpus}
  \Description{EDP versus frequencies and powercap settings for synthetic benchmarks and Gromacs on A40 and A100 GPUs. For Figure (a) with the synthetic benchmarks, the EDP is normalized.}
\end{figure*}

\autoref{fig:edpxfreq_gpu} displays EDP versus fixed frequency for both SGEMM and Triad benchmarks, and \autoref{fig:edpxfreq_gpu} displays the same plot for the Gromacs benchmarks.
Both cases with a high computational workload reveal similar EDP behavior.
There is a significant improvement from the minimum frequency to the base frequency.
Then, the EDP starts to increase slightly, showing that the relationship between performance and energy consumption remains nearly linear.
\autoref{fig:edpxpowercap_gpu} shows the relative EDP versus power capping settings with Gromacs.
The EDP behavior is similar at different scales and for the frequency setting, but the EDP improvement is relatively smaller for the A100 GPU compared to the A40.
Since EDP is normalized, this effect is likely due to better EDP measurements with the lowest power capping setting for the A100 and for the B4 (which has a smaller workload than the B5) for the A40.

These experiments and discussions make it clear that studies on energy and performance analysis must consider several factors related to the application and the hardware on which it runs.
It is important to avoid overgeneralized claims and remember that results may be limited to the utilized benchmark and hardware.

\subsection{Affinity control}

Affinity control, which restricts which physical cores, threads and tasks can run on, is critical for meaningful energy measurements.
On some CPUs, energy can only be measured at socket level. Since the frequency of the cores depends on how many are active, it is important to use all the cores on a socket for energy measurements.
In modern HPC systems, tools and runtime environments influence or restrict affinity settings at several levels.
The SLURM scheduler can either put the user context into a restricted OS CPU set, or, when used as a startup mechanism for an MPI or hybrid MPI+OpenMP application, it can enforce affinity settings.
Common MPI startup tools such as OpenMPI's mpirun and Intel MPI's mpiexec also support proprietary affinity control.
Some profiling tools, such as the Likwid tools, also provide affinity control.
Finally, the application itself can control its affinity. 
When it comes to benchmarking, it is important that the available tools are utilized properly and do not interfere with each other.
Often, a dedicated test program that only determines and prints the actual affinity settings is required to check that the settings are working as expected and debug any issues.

For instance, in our experiments with Gromacs there are three possible layers of pinning/CPU binding: (a) SLURM, which limits the available amount of CPUs according to its setting, (b) Likwid, which pins threads and properly bind ranks to the available resources in the compute nodes, and (c) Gromacs, which has its own default pinning mechanisms for performance purposes.

In \autoref{fig:b4_freqxpower_cpu_incorrect_pinning} we illustrate the importance to ensure that ranks are properly mapped to a single socket.
As shown it is possible that inconsistent pinning led to strange behaviors like the CPU frequency on SPR operating at higher core frequencies than the expected average when using the full socket (about 3.6GHz in some cases, where core frequencies usually should not exceed 2.6GHz).
The reason for these higher frequencies is that since the socket was not fully utilized, the CPU cores could clock higher and therefore reach better performance even though the same number of tasks was used.
The fact that the average power draw plateaus at the TDP value for a single socket makes the issue harder to detect.

To mitigate this, SLURM is configured to restrict the available CPU cores to the ones in the same socket and then execute  \textbf{likwid-mpirun} wrapped up around the \textbf{srun} command, meaning that the only options for Likwid to pin are on the restricted sockets.
Since Gromacs is able to identify external pinning and then refrain from actually perform the pinning, it did not interfere with the external affinity settings.
It still might be important to evaluate different pinning strategies, since Gromacs can perform pinning using application knowledge and even dynamically change pinning behavior during program execution.


\subsection{Control over chip settings}

\begin{figure*}[tb]
  \centering
  \begin{subfigure}{0.33\textwidth}
    \includegraphics[width=\linewidth]{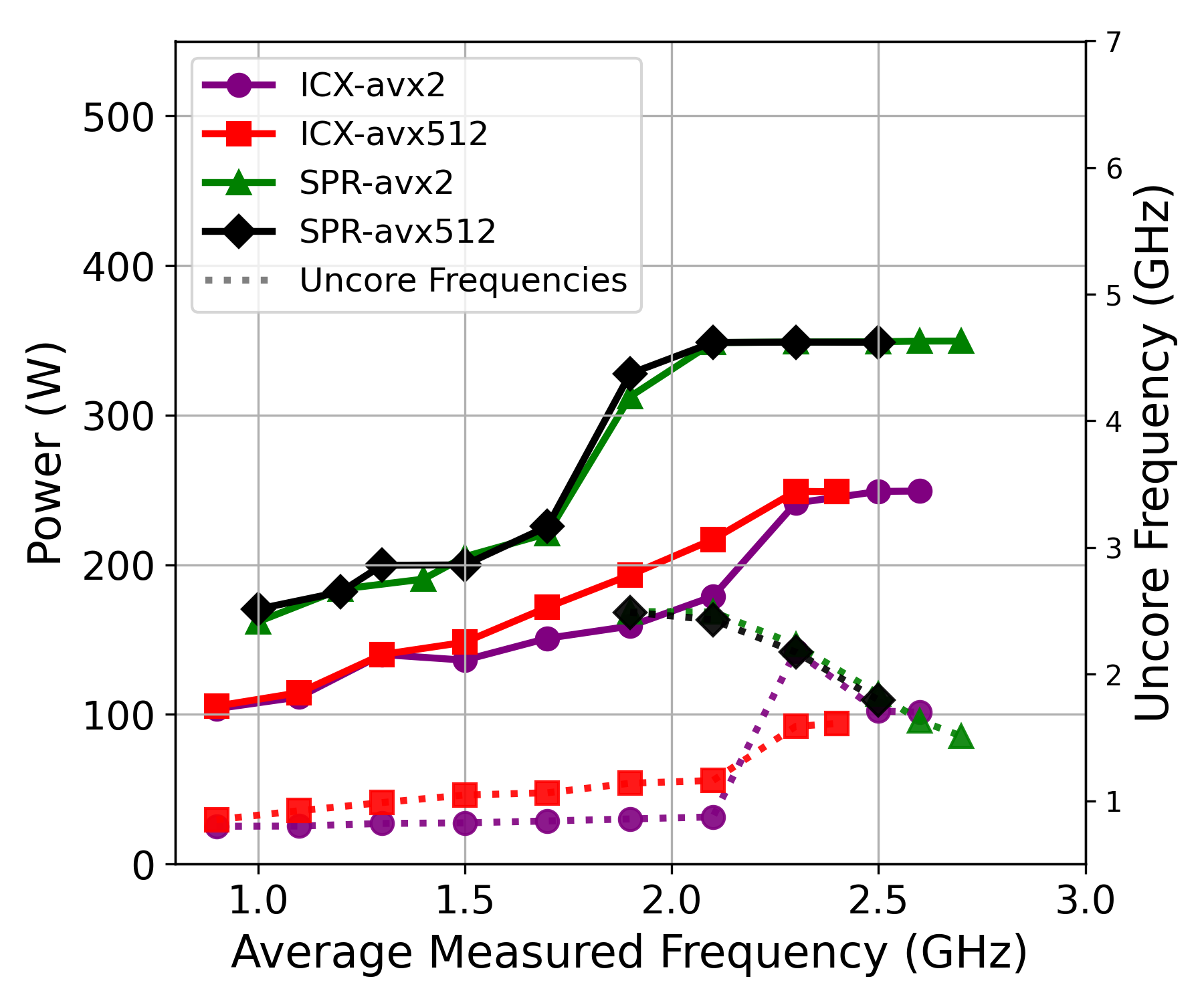}
    \caption{Gromacs B4 on CPUs}
    \label{fig:b4_freqxpower_cpu}
  \end{subfigure}
  \begin{subfigure}{0.33\textwidth}
    \includegraphics[width=\linewidth]{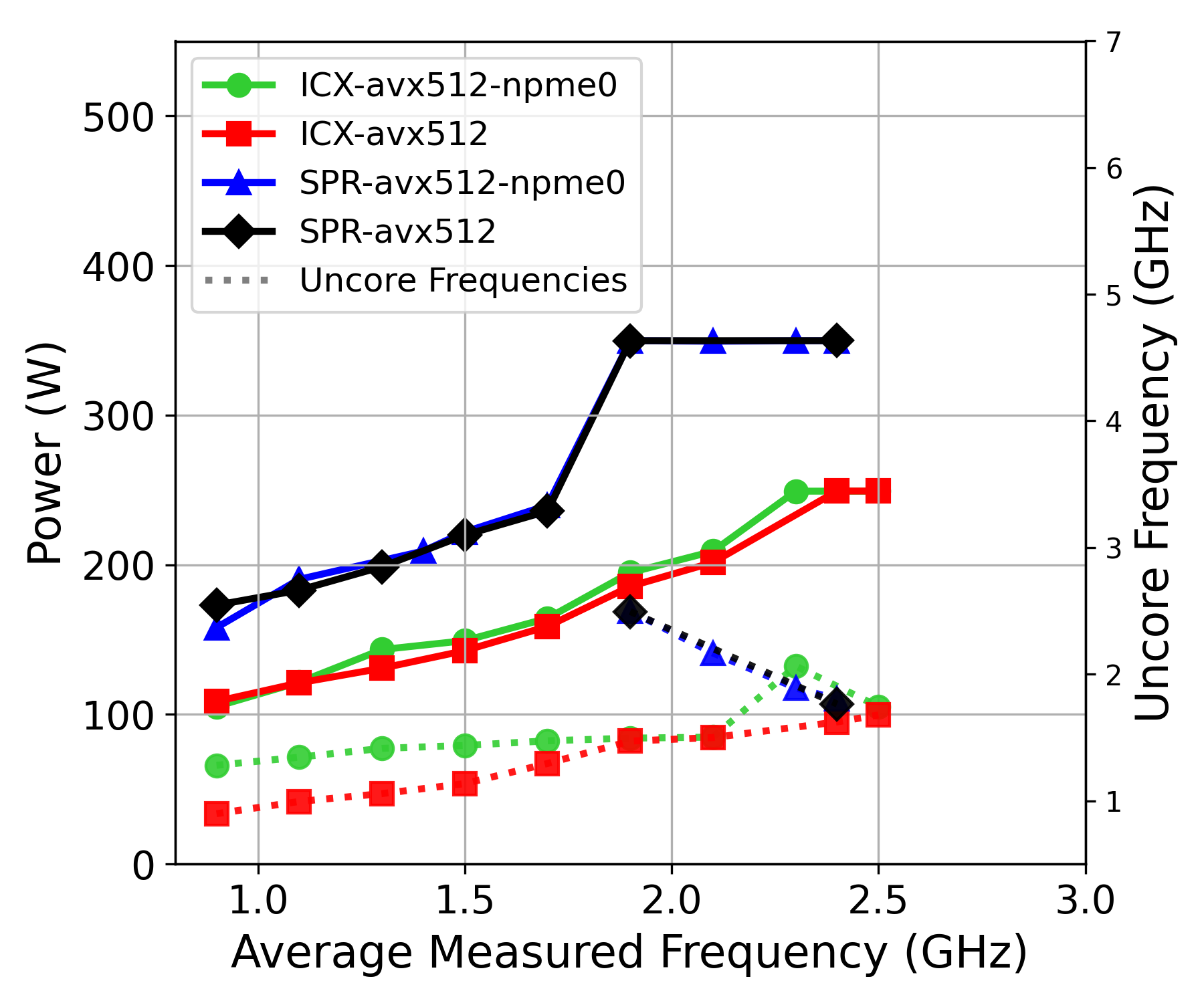}
    \caption{Gromacs B5 on CPUs}
    \label{fig:b5_freqxpower_cpu}
  \end{subfigure}
  \begin{subfigure}{0.33\textwidth}
    \includegraphics[width=\linewidth]{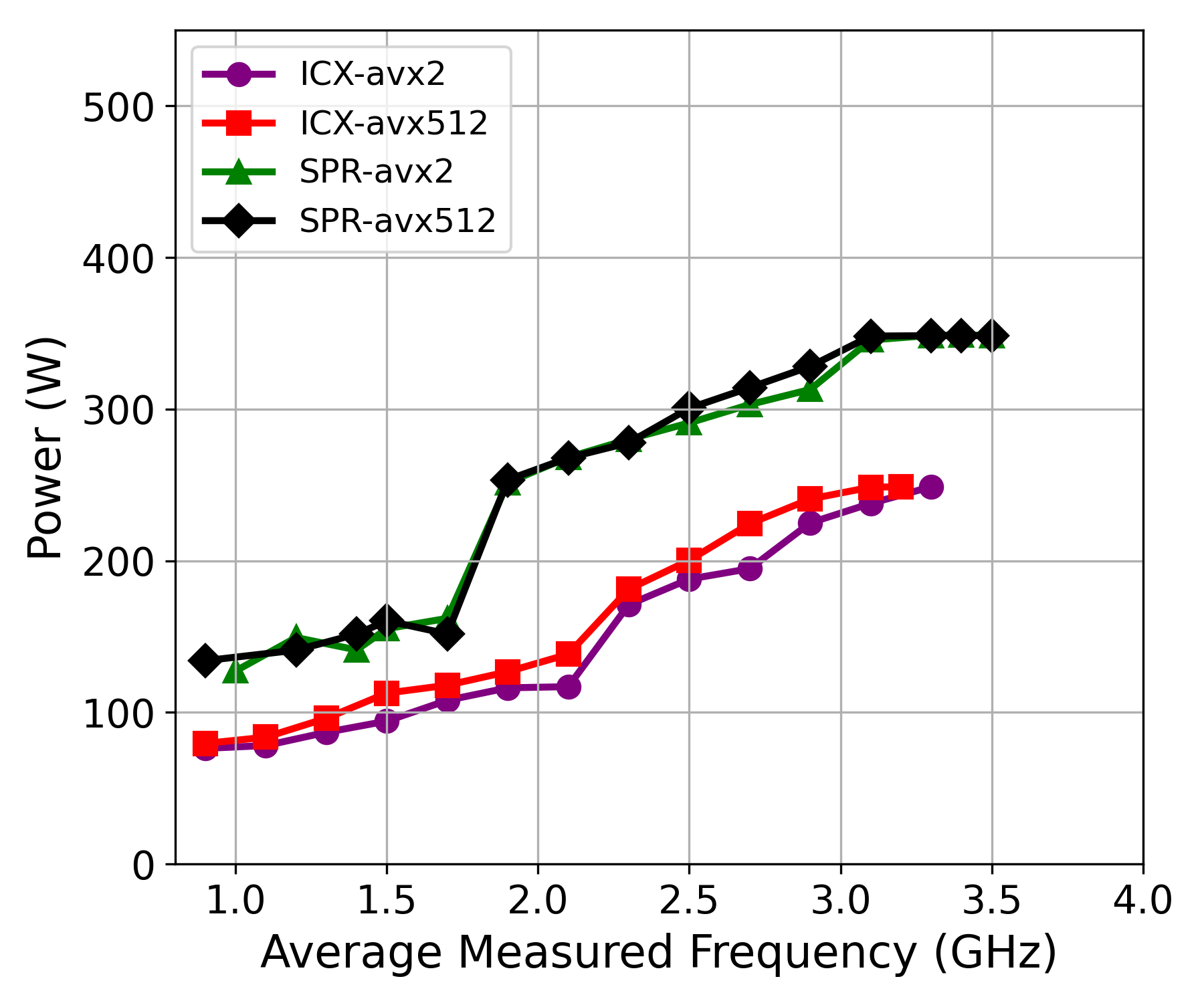}
    \caption{Gromacs B4 on CPUs, incorrect pinning}
    \label{fig:b4_freqxpower_cpu_incorrect_pinning}
  \end{subfigure}
  \caption{Average power draw versus measured frequencies for Gromacs on CPUs. Dotted lines show the measurements from uncore frequencies for each case. Figure (c) shows the measurements for cases in which the ranks are distributed between the two sockets.}
  \label{fig:freqxpowers}
  \Description{Average power draw versus measured frequencies for Gromacs on CPUs. Dotted lines show the measurements from uncore frequencies for each case. Figure (c) shows the measurements for cases in which the ranks are distributed between the two sockets.}
\end{figure*}

For energy studies, the ability to set power cap limits or explicitly enforce frequencies is essential.
Unfortunately, the dynamic frequency scaling behavior of modern CPUs and GPUs means that these settings are treated as mere suggestions by the processor.
Even the base frequency is not guaranteed on Intel CPUs when running hot SIMD instruction code.
In studies, it is also important to include the Turbo frequency range, as this often comprises more than a third of the available frequency range.
Even in situations where a specific behavior is expected for a certain parameter, it is important to confirm this via measurements and to use the measurements instead of the expected setting, if possible.
However, it is not uncommon for studies to show the set frequency rather than the actual measured values.

\autoref{fig:freqxpowers} shows an example where displaying the average frequency can be elucidative: starting from the base frequencies of both architectures, the power draw already reaches and remains close to the TDP.
However, we can see that the measured frequency increases slightly when the frequency setting is increased, even though the power remains roughly constant.
This is because the uncore frequency of the CPU socket decreases slightly, providing some room for the CPU cores to clock higher.
Consequently, the energy and power consumption are moderately shifted towards the cores when the frequency setting is higher.
The energy-to-solution is decreased in both cases as can be seen in \autoref{fig:b4_perfxenergy_cpu} and \autoref{fig:b5_freqxpower_cpu} corroborates that this actually provides some marginal benefits to both performance and energy consumption, which makes evident that for these Gromacs benchmarks the bottleneck is the CPU cores and not the other domains.

If only the specified frequency were shown in this example, it would not be possible to perceive this effect because one could simply assume that the frequency remains fixed during the final executions in turbo mode.
While one could make this assumption based on the shorter runtime, it would still not be evident because the runtime is influenced by many other factors.
Therefore, in most cases, it is more interesting to show the actual frequency measurement than the setting itself because any divergence between the two makes the latter deceptive.
Note that measurements can also be misleading or even inaccurate in some cases, either due to the previously discussed overhead situation or simply because the measurement tool is unreliable.
To mitigate this, it is recommended that measurements are validated against benchmarks where their values can be estimated, such as synthetic benchmarks, and that they are cross-validated with other measurement tools.


\subsection{Benchmarking methodology}

A systematic and consistent benchmarking methodology is essential to achieve meaningful and predictable results.
Most important is to have a clear idea of the expected results.
Without this, subtle errors and incorrect measurements cannot be identified.
This requires experience; if none is available, it is recommended that one asks other researchers for advice or find comparable results from others.
More experiments using different tools or settings need to be conducted, particularly when the results are too good or bad, or if the shape of a plot contradicts the expectation, to build trust in the result.
When a result is unexpected, it can be difficult to find an explanation for it.
Additional experiments can either provide an explanation or reveal errors in previous measurements.
Although an automated benchmarking script using batch jobs is convenient, consistent, and allows more data to be gathered, there are reasons to do benchmarking manually in an interactive session first.
With automated measurements, there is a tendency to acquire more data than is required.
Having too much data makes it more difficult to identify errors, since it is usually no longer possible to examine all of the raw results.
While automated scripts are convenient, the turnaround time is often much slower since running all benchmarks can take hours.
Although doing all steps by hand might seem tedious, issues are detected faster and more easily since they are not buried in complicated scripts, but rather occur right before your eyes.
There is also direct feedback if something takes longer than expected. For example, one might use this feedback to detect a large profiling overhead.
In this case, fixing and repeating the test case is a matter of minutes, and the turnaround time is much faster, especially if multiple attempts are necessary to get it right.

\section{Conclusion}
\label{sec:conclusion}

This paper discussed the challenges and pitfalls of conducting energy efficiency studies on real application codes for modern high-performance computing (HPC) systems.
It showed measurements for Gromacs and MD-Bench, a molecular dynamics prototyping harness, as well as multiple synthetic benchmarks for CPUs and GPUs.
Meaningful energy measurements are difficult to produce due to the dynamic frequency scaling of modern processors, which depends on the number of utilized cores, code characteristics, processor specimens, and server environmental conditions.
Mitigating this issue requires complete control of affinity settings and measurement and displaying the average frequency rather than the set properties.
Due to the high overhead and low sampling frequencies of energy counters, as well as the general lack of support for HPM metrics, a thorough measurement strategy is necessary for accurate results.
While RAPL counters enabled much energy-related research, it would be desirable to query additional data sources, such as on-board management chips or PSUs, more easily.
The same applies to GPUs, for which measurements can only be performed using the tools and drivers of GPU vendors, which cannot be verified further.
In this case, too, a standardized, open interface on the motherboard for GPU power draw would be beneficial.
Due to the difficult environment and the many sources of error, it is recommended to not rely solely on an automated benchmark script since errors are difficult to detect in this case.
In this context, developing a clear result expectation is important for validating and challenging the measured results.

\begin{acks}
This work was funded by the German Federal Ministry of Research, Technology, and Space (BMFTR) under the funding program Hoch- und Höchstleistungsrechnen für das digitale Zeitalter 2021-2024 – Forschung und Investitionen zum High-Performance Computing, Call Richtlinie zur Förderung von Verbundprojekten auf dem Gebiet des energieeffizienten High-Performance Computings (GreenHPC) as part of the EE-HPC project (Grant 16ME0583K).
\end{acks}

\bibliographystyle{ACM-Reference-Format}
\bibliography{references}

\appendix

\end{document}

%% file: macros.tex
\newcommand{\bq}{\begin{equation}}
\newcommand{\eq}{\end{equation}}
\newcommand{\bytes}{\mbox{bytes}}
\newcommand{\byte}{\mbox{byte}}
\newcommand{\second}{\mbox{s}}
\newcommand{\seconds}{\mbox{s}}
\newcommand{\flop}{\mbox{flop}}
\newcommand{\flops}{\mbox{flops}}
\newcommand{\NJFLOP}{\mbox{nJ/\flop}}
\newcommand{\instr}{\mbox{instr}}
\newcommand{\cycle}{\mbox{cy}}
\newcommand{\iter}{\mbox{it}}
\newcommand{\cycles}{\mbox{cy}}
\newcommand{\FCY}{\mbox{\flop/\cycle}}
\newcommand{\FIT}{\mbox{\flop/\iter}}
\newcommand{\BIT}{\mbox{\byte/\iter}}
\newcommand{\FR}{\mbox{\flops/\mbox{row}}}
\newcommand{\BR}{\mbox{\byte/\mbox{row}}}
\newcommand{\CYF}{\mbox{\cycles/\flop}}
\newcommand{\CS}{\mbox{\cycle/\second}}
\newcommand{\GCS}{\mbox{G\cycle/\second}}
\newcommand{\word}{\mbox{Word}}
\newcommand{\words}{\mbox{Words}}
\newcommand{\order}[1]{\mbox{${\cal O}\left(\mbox{#1}\right)$}}
\newcommand{\bit}{\mbox{bit}}
\newcommand{\bits}{\mbox{bits}}
\newcommand{\GBPS}{\mbox{G\bit/\second}}
\newcommand{\MBPS}{\mbox{M\bit/\second}}
\newcommand{\FS}{\mbox{\flop/\second}}
\newcommand{\BS}{\mbox{\byte/\second}}
\newcommand{\BC}{\mbox{\byte/\cycle}}
\newcommand{\GBS}{\mbox{G\byte/\second}}
\newcommand{\MBS}{\mbox{M\byte/\second}}
\newcommand{\GWS}{\mbox{G\word/\second}}
\newcommand{\GFS}{\mbox{G\flop/\second}}
\newcommand{\MFS}{\mbox{M\flop/\second}}
\newcommand{\lup}{\mbox{LUP}}
\newcommand{\lups}{\mbox{LUPs}}
\newcommand{\LUPCY}{\mbox{\lup/\cycle}}
\newcommand{\LUPS}{\mbox{\lup/\second}}
\newcommand{\MLUPS}{\mbox{M\lup/\second}}
\newcommand{\GLUPS}{\mbox{G\lup/\second}}
\newcommand{\GHZ}{\mbox{GHz}}
\newcommand{\ns}{\mbox{ns}}
\newcommand{\WF}{\mbox{\word/\flop}}
\newcommand{\BF}{\mbox{\byte/\flop}}
\newcommand{\FB}{\mbox{\flop/\byte}}
\newcommand{\BL}{\mbox{\byte/\lup}}
\newcommand{\GB}{\mbox{GB}}
\newcommand{\KB}{\mbox{kB}}
\newcommand{\MB}{\mbox{MB}}
\newcommand{\GiB}{\mbox{GiB}}
\newcommand{\MiB}{\mbox{MiB}}
\newcommand{\KiB}{\mbox{KiB}}
\newcommand{\W}{\mbox{W}}
\newcommand{\muarch}{\mbox{$\mu$-arch}}
\newcommand{\muop}{\mbox{$\mu$-op}}
\newcommand{\muops}{\mbox{$\mu$-ops}}
\newcommand{\eos}{\;.}
\newcommand{\cma}{\;,}
\newcommand{\rlm}{roof{}line model}
\newcommand{\rl}{roof{}line}
\newcommand{\Rlm}{Roof{}line model}
\newcommand{\Rl}{Roof{}line}
\newcommand{\olsep}{\|}
\newcommand{\nolsep}{|}
\newcommand{\ecmspace}{\,}
\newcommand{\TOL}{$T_{\mathrm{c}\_\mathrm{OL}}$}
\newcommand{\NNZR}{$N_\mathrm{nzr}$}
\newcommand{\NR}{$N_\mathrm{r}$}
\newcommand{\NNZ}{$N_\mathrm{nz}$}
\newcommand{\ecm}[6]{\mbox{$\left\{{#1}\ecmspace\olsep\ecmspace {#2}\ecmspace\nolsep\ecmspace {#3}\ecmspace\nolsep\ecmspace {#4}\ecmspace\nolsep\ecmspace {#5}\right\}\ecmspace{#6}$}}
\newcommand{\epsep}{\rceil}
\newcommand{\ecmp}[4]{\mbox{$\left\{{#1}\ecmspace\epsep\ecmspace {#2}\ecmspace\epsep\ecmspace {#3}\right\}\ecmspace{#4}$}}
\newcommand{\ecme}[4]{\mbox{$\left({#1}\ecmspace\epsep\ecmspace {#2}\ecmspace\epsep\ecmspace {#3}\right)\ecmspace{#4}$}}
\newcommand{\sellcs}{SELL-\texorpdfstring{$C$-$\sigma$}{C-sigma}}
\newcommand{\likwid}{\texttt{LIKWID}}
\newcommand{\likwidperfctr}{\texttt{likwid-perfctr}}
\newcommand{\likwidpin}{\texttt{likwid-pin}}
\newcommand{\likwidbench}{\texttt{likwid-bench}}
\newcommand{\lmbench}{\texttt{lmbench}}
\definecolor{tumbleweed}{rgb}{0.87, 0.67, 0.53}
\newcommand{\afx}{A64FX}
\newcommand{\spmv}{SpMV}
\newcommand{\cmg}{CMG}
\newcommand{\mve}{MVE}
\newcommand{\crs}{CRS}
\newcommand{\ellpack}{ELLPACK}
\newcommand{\tands}{dRECT}
\newcommand{\Fig}{Figure}
\newcommand{\Figure}{Figure}
\newcommand{\Figures}{Figures}
\newcommand{\Sect}{Section}
\newcommand{\Sects}{Sections}
\newcommand{\Section}{Section}
\newcommand{\Sections}{Sections}
\newcommand{\DMVM}{\textsc{DMVM}}

\newcommand{\mathspace}{\text{ }}
\newcommand{\rAdd}[1]{{\color{red}{#1}\color{black}}}


\newcommand{\inputtikz}[2]{%
	\input{#1/#2.tex}%
}

\newenvironment{customlegend}[1][]{%
	\begingroup
	\csname pgfplots@init@cleared@structures\endcsname
	\pgfplotsset{#1}%
}{%
	\csname pgfplots@createlegend\endcsname
	\endgroup
}%
\def\addlegendimage{\csname pgfplots@addlegendimage\endcsname}

\definecolor{applegreen}{rgb}{0.55, 0.71, 0.0}
\definecolor{amethyst}{rgb}{0.6, 0.4, 0.8}
\definecolor{amber}{rgb}{1.0, 0.75, 0.0}